\documentclass[%
  reprint,
  superscriptaddress,
  amsmath,amssymb,
  aps,
  pre 
]{revtex4-2}

\usepackage[T1]{fontenc}
\usepackage[utf8]{inputenc}
\usepackage{bm}
\usepackage{physics}
\usepackage{graphicx}
\usepackage{xcolor}
\usepackage{booktabs}      
\usepackage{hyperref}
\hypersetup{
  colorlinks=true,
  linkcolor=blue!50!black,
  citecolor=teal!50!black,
  urlcolor=magenta!60!black
}
\usepackage[nameinlink]{cleveref}
\allowdisplaybreaks[2]

\newcommand{\E}{\mathbb{E}}
\newcommand{\Var}{\mathrm{Var}}
\newcommand{\RR}{\mathbb{R}}
\newcommand{\PP}{\mathbb{P}}
\newcommand{\1}{\mathbf{1}}

\begin{document}

\title{First Passage Problem:\\ Asymptotic Corrections due to Discrete Sampling }

\author{Lars Fritz}
\affiliation{Institute for Theoretical Physics, Utrecht University,
Princetonplein 5, 3584 CC Utrecht, The Netherlands}
\email{l.fritz@uu.nl}

\date{\today}

\begin{abstract}
How long a stochastic process survives before leaving a domain depends
not only on its intrinsic dynamics but also on \emph{how it is observed}.
Classical first-passage theory assumes continuous monitoring with
absorbing boundaries (``kill-on-touch''). In practice, however,
measurements are often taken at discrete times. Between two checks, a
trajectory may leave and re-enter the domain without being detected.
Under this \emph{stroboscopic} rule (``kill-on-check''), exit statistics
change qualitatively.

We analyze one-dimensional Brownian motion confined to an interval of
length $L$ and observed at frame intervals~$\Delta t$, with diffusive
step scale $\sigma\sqrt{\Delta t}$. The dynamics collapse onto a single
confinement ratio $\rho=L/(\sigma\sqrt{\Delta t})$. For boundary starts
we obtain linear scaling of the mean number of frames until
exit, while for bulk starts the survival is governed by the spectral gap
of a one-step stroboscopic operator, leading to a quadratic law with linear corrections. These results identify the stroboscopic
first-passage problem where the observation protocol itself reshapes the statistics of
escape.
\end{abstract}

\maketitle

\section{Introduction}
\label{sec:intro}

The question of how long a random process remains confined before it
escapes a domain is a cornerstone of probability theory and statistical
physics. From chemical reactions and neuronal firing to diffusion in
cells and finance, \emph{first-passage problems} quantify how
fluctuations meet boundaries. Traditionally, such problems are cast in a
continuous-monitoring framework: a Brownian particle is absorbed
immediately upon touching a boundary, implemented via Dirichlet boundary
conditions. This ``kill-on-touch'' idealization underlies much of
classical first-passage theory
\cite{Redner,MetzlerBook,BorodinSalminen}.

\begin{figure}[t]
  \centering
  \includegraphics[width=0.9\linewidth]{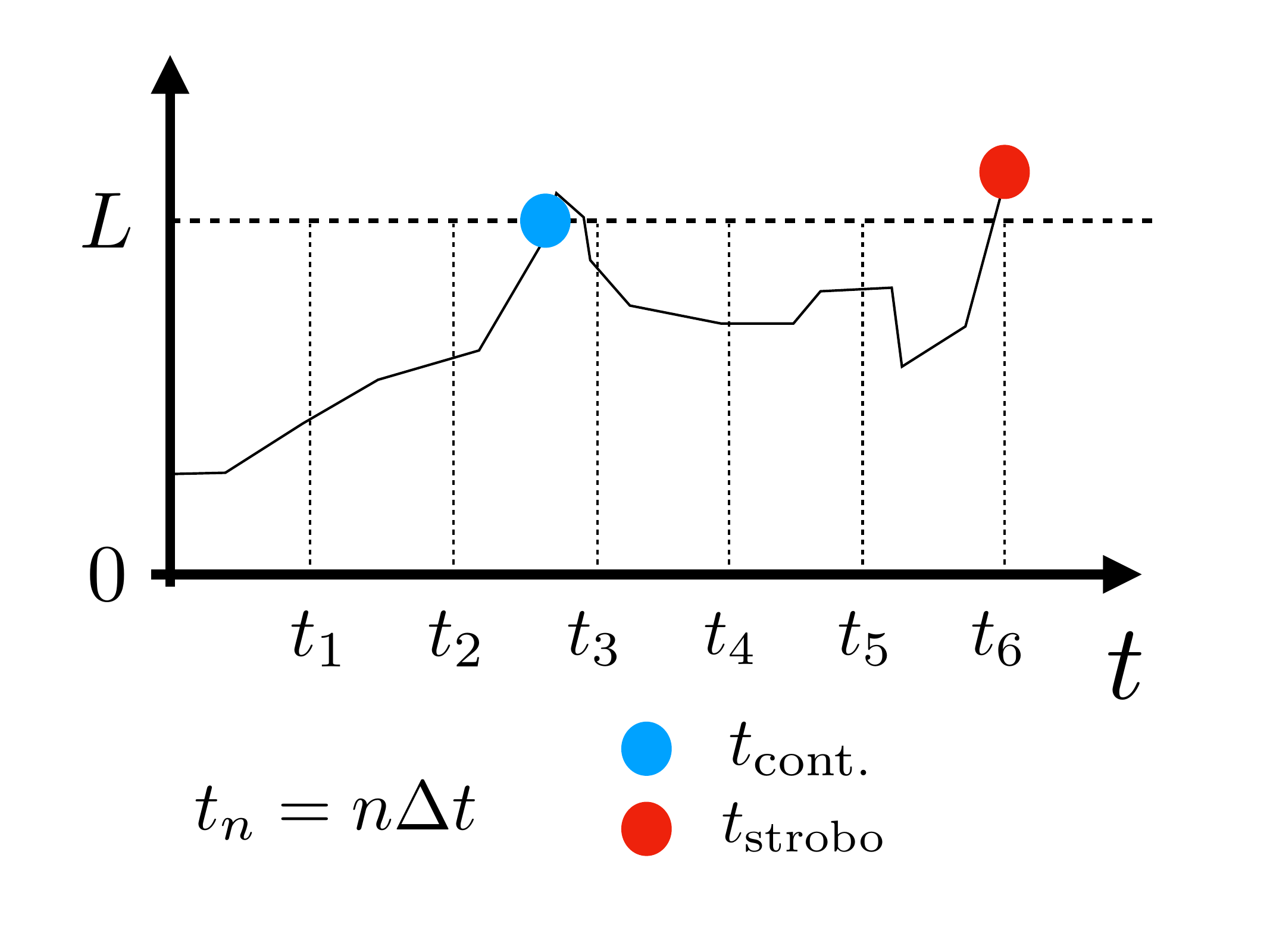}
  \caption{
  Schematic of stroboscopic monitoring. A Brownian trajectory (gray line) is sampled only at discrete observation times $t_n = n\Delta t$ (vertical markers). Between frames the motion is continuous, allowing \emph{undetected excursions}—segments that leave and re-enter the confining domain before the next check. Under this ``kill-on-check'' rule, absorption is enforced only at sampling instants, in contrast to continuous (Dirichlet) monitoring where escape is detected immediately.
  }
  \label{fig:strobot}
\end{figure}

In practice, however, measurement and sampling are often discrete.
Imaging systems, numerical samplers, or other detectors record positions
only at time intervals of duration~$\Delta t$, which in some cases may
vary. Between two such frames, the process can exit and re-enter the
confinement region without being observed—an \emph{undetected excursion}, see Fig.~\ref{fig:strobot}.

Under such \emph{stroboscopic observation}, absorption is enforced only
at sampling times, effectively replacing continuous absorption by a
``kill-on-check'' rule. Remarkably, this seemingly minor modification of
the observation protocol qualitatively transforms survival statistics.

To study this effect in a controlled setting, we focus on one of the
simplest nontrivial examples: one-dimensional Brownian motion (diffusion
coefficient~$D=\sigma^2/2$) confined to an interval~$(0,L)$ and observed
at discrete times~$t_n=n\Delta t$. The motion itself remains continuous
between frames, but exits are registered only at sampling instants. The
dynamics depend on a single dimensionless parameter,
\[
\rho = \frac{L}{\sigma\sqrt{\Delta t}},
\]
which compares the confinement length~$L$ to the typical diffusive
displacement per frame.

In this paper, we introduce a compact and flexible theoretical framework
based on a path-integral evolution: each frame consists of free Gaussian
propagation over~$\Delta t$ followed by projection onto the confined
region. This yields the one-step operator
\[
K = P\,G_0\,P,
\]
where $G_0$ is the free propagator and $P$ the projector onto~$(0,1)$
(after rescaling). In this construction, discrete sampling enters at the
operator level rather than as an external approximation. The framework
supports analytic asymptotics, spectral and resolvent analysis,
deterministic Nyström discretization, and extensions to include drift,
random frame intervals, or asymmetric domains.

Within this framework we derive modifications to classical
first-passage behavior.  
Let $\tau$ denote the first frame index at which the particle is found
outside the interval, i.e.\ the discrete first-exit time (measured in
number of frames).  
For a boundary start, continuous absorption would imply immediate exit,
yet under stroboscopic monitoring we obtain
\[
\E[\tau](\rho;0)
  = \frac{\rho}{\sqrt{2}}
  + \frac{|\zeta(\tfrac12)|}{\sqrt{\pi}} + \ldots ,
\]
so that the mean number of frames until exit grows linearly with~$\rho$.
For a bulk start (e.g.\ $y_0=\tfrac12$), survival is governed by the
spectral gap of~$K$, yielding the diffusive leading law together with subleading corrections,
\[
\E[\tau](\rho;\tfrac12)
  = \tfrac14\,\rho^2 + 0.5830\,\rho + 0.5736 + \ldots .
\]
The linear and constant terms are large enough in realistic parameter
ranges to alter the apparent scaling of survival times, mimicking
anomalous exponents over finite windows. This transient bias may help rationalize observations of apparent subdiffusion
in discretely sampled trajectories, a question of direct experimental relevance. Their structure matches the
exact asymptotics derived by Lotov~\cite{Lotov96} for Gaussian random
walks between two barriers, where the remainder is exponentially small
in~$\rho$. We provide an independent rederivation of these results in a
more physics-oriented language based on the projector–resolvent
formalism and find that the resulting constants coincide precisely with
Lotov’s values and also confirm them with our efficient numerical scheme.

These results go beyond refinements of the leading theory: they reveal
that discrete observation fundamentally reshapes first-passage behavior.
Since the corrections are built into the operator-level formulation,
they are an intrinsic component of the dynamics under stroboscopic
monitoring rather than a small perturbative effect.

The novel operator framing also allows straightforward generalizations.
In particular, we extend the analysis to random sampling intervals.  Let
$\mu(u)$ denote the probability density of frame durations $u>0$ with
mean $\overline{\Delta t}=\int u\,\mu(u)\,du$.  For a given distribution
of sampling intervals, one defines an averaged one-step operator
\[
K_\mu = P \Bigl(\int_0^\infty G_u\,\mu(u)\,du\Bigr)\,P,
\]
where $G_u$ is the free Gaussian propagator for a time step of length
$u$.  This construction preserves the leading slope in the
boundary-start regime but shifts the subleading constants according to
the variance of the sampling times, $\Var(U)$.  Among all sampling laws
with the same mean, perfectly regular (equal) spacing minimizes the
additive constant.

\textbf{Outline.} The rest of the paper is organized as follows. 
In Sec.~\ref{sec:basics} we revisit continuous diffusion and its three equivalent formulations (PDE, Green function, Wiener integral). 
Sec.~\ref{sec:survival} contrasts continuous and stroboscopic survival and defines the discrete exit time~$\tau$. 
Sec.~\ref{sec:dirichlet} presents the continuous Dirichlet benchmark via the resolvent. 
Sec.~\ref{sec:strobo-framework} constructs the one-step operator and projector–resolvent framework. 
Sec.~\ref{sec:asymptotics} states and derives the boundary and bulk asymptotic laws. 
Sec.~\ref{sec:random} treats random frame times via propagator averaging. 
Sec.~\ref{sec:numerics} details the Nyström implementation and confirms the asymptotics numerically. 
We conclude in Sec.~\ref{sec:outlook}.

In contrast to earlier approaches, our method does not rely on
\emph{post hoc} ``continuity corrections'' or barrier shifts to adapt
continuous-time formulas to discrete sampling. For instance, Broadie,
Glasserman, and Kou proposed an empirical barrier adjustment
$\exp(\beta\sigma\sqrt{\Delta t})$ (with $\beta\!\approx\!0.5826$) to
account for discrete monitoring in barrier options
\cite{BroadieGlassermanKou97}, while matched-asymptotic expansions were
later developed to generalize this idea to broader monitoring schemes
\cite{HowisonStein05}. In financial mathematics, Li and Linetsky
introduced eigenfunction-expansion techniques that treat discrete
monitoring as a recursive correction to the continuous problem
\cite{LiLinetsky15}. All of these approaches, however, modify the
continuum theory \emph{after the fact}: they do not embed the
discrete-sampling mechanism at the operator or path-integral level, and
they target pricing applications rather than first-passage
asymptotics. Our framework, by contrast, builds the discrete observation
directly into the evolution operator itself, yielding both the
boundary-start linear law and the bulk-start correction as intrinsic
features of the dynamics—analogous in spirit to how Nezlobin and Tassy
incorporate deterministic block timing in their analytical LVR model
\cite{TassyNezlobin25}.

\paragraph*{Broader context.}
The framework developed here is general and lends itself to applications
well beyond the purely theoretical setting considered below.  
In particular, it provides the foundation for quantitative analyses of
discrete-sampling effects in experimental single-particle tracking and
for analogous phenomena in systems with stochastic timing such as
block-chain transaction flows or asynchronous numerical schemes.
Those extensions will be explored elsewhere.
\section{The setup: Diffusion equation, the Green function, and the Wiener path integral}
\label{sec:basics}

To prepare for the analysis of stroboscopic monitoring, we briefly review the continuum formulation of Brownian motion and its standard mathematical representations. The diffusion equation, its Green function, and the Wiener path integral provide three equivalent viewpoints: a local partial differential equation for the evolution of probability, an integral kernel (propagator) that solves it, and a trajectory-based functional integral expressing its additivity. These elements form the reference framework for all subsequent extensions in which observation is discrete and boundaries are enforced only at sampling times.

\subsection{Diffusion equation as conservation law}
Let $p(x,t)$ denote the probability density for a Brownian particle on the real line.  
Its evolution obeys the \emph{diffusion equation}
\begin{equation}
\label{eq:diffusion}
\partial_t p(x,t) = D\,\partial_x^2 p(x,t),
\qquad D=\frac{\sigma^2}{2},
\end{equation}
which expresses local conservation of probability in the form 
$\partial_t p + \partial_x J = 0$, with diffusive flux 
$J=-D\,\partial_x p$.  
This equation arises as the continuum limit of a random walk with step variance $\sigma^2 \Delta t$ and becomes exact for Brownian motion in the limit of infinitesimal steps.  
Equation~\eqref{eq:diffusion} is linear, Markovian, and self-adjoint, ensuring conservation of total probability and positivity of the density for all $t>0$.

\subsection{Green function and free propagation}
The fundamental solution of the diffusion equation is the response to a point source at $x_0$, 
\begin{eqnarray}
\partial_t G_0(x,t|x_0) &=& D\,\partial_x^2 G_0(x,t|x_0)\;,\nonumber \\ 
G_0(x,0|x_0)&=&\delta(x-x_0)\;.
\end{eqnarray}
Its explicit form,
\begin{equation}
\label{eq:G0}
G_0(x,t|x_0)
=\frac{1}{\sqrt{4\pi D t}}
\exp\!\left[-\frac{(x-x_0)^2}{4Dt}\right],
\end{equation}
is the Gaussian heat kernel.  
It represents the probability density for a particle initially at $x_0$ to be found at $x$ after time $t$.  
The kernel is normalized, $\int_{\RR} G_0(x,t|x_0)\,\dd x=1$, reflecting particle conservation, and satisfies the \emph{semigroup property}
\begin{equation}
\int_{\RR} G_0(x,t_2|x_1)\,G_0(x_1,t_1|x_0)\,\dd x_1
   =G_0(x,t_1{+}t_2|x_0),
\end{equation}
which expresses the additive evolution of a Markov process in time. This
Markov (memoryless) property will later underlie the construction of the
stroboscopic propagator as a sequence of Gaussian steps interleaved with
projections.

\subsection{Spectral representation and mode filtering}
Since the Laplacian $\partial_x^2$ is diagonal in plane waves $e^{ikx}$, one may equivalently write
\[
G_0(x,t|x_0)
=\int_{\RR}\frac{dk}{2\pi}\,e^{ik(x-x_0)}\,e^{-Dk^2t}.
\]
Diffusion thus acts as a Gaussian filter in Fourier space, damping each mode with rate $Dk^2$.  
This spectral viewpoint will be useful when projectors are later introduced in $k$–space, as it makes explicit how discrete observation selectively suppresses or reweights spatial modes.

\subsection{Wiener path integral and additivity of the action}
The same propagator can be obtained from a sum over all continuous trajectories connecting $x_0$ and $x$ in time $t$.  
Dividing $t$ into $N$ equal subintervals of length $\varepsilon=t/N$ and using the Markov property, the joint transition density factorizes:
\[
p(x_N,t|x_0)=\!\!\int\!\dd x_1\!\cdots\!\dd x_{N-1}
   \prod_{j=1}^{N} G_0(x_j,\varepsilon|x_{j-1})
\]
where $x_N=x$.
In the continuum limit this product becomes the Wiener path integral
\begin{equation}
G_0(x,t|x_0)
=\!\!\int_{x(0)=x_0}^{x(t)=x}
\!\!\exp\!\left[-\frac{1}{2\sigma^2}
      \int_0^t \dot x(s)^2\,\dd s\right]\!
\mathcal{D}x
\end{equation}
with $\sigma^2=2D$.
The exponent is \emph{additive} under concatenation of paths:
for any intermediate time $t'=t_1+t_2$,
\[
\int_0^{t'}\!\dot x^2\,\dd s
= \int_0^{t_1}\!\dot x^2\,\dd s
+ \int_{t_1}^{t'}\!\dot x^2\,\dd s,
\]
which ensures the semigroup property of $G_0$ and establishes the connection between the path-integral and differential formulations.  
This additive structure will later carry over directly to the stroboscopic case, where each time slice is terminated by a projection gate rather than a simple concatenation.

\subsection{Complementarity and outlook}
The diffusion equation describes the local evolution law, the Green function its global propagator, and the Wiener integral its trajectory-wise representation.  
Together they provide the analytic and probabilistic foundation for free Brownian motion.  
In the following sections, boundaries and discrete monitoring will be introduced, turning continuous propagation into a sequence of alternating operations—free Gaussian spreading for one frame and projection back onto the domain.  
It is this alternation that fundamentally reshapes survival statistics in the stroboscopic first-passage problem.

\section{Survival: continuous vs.\ stroboscopic}
\label{sec:survival}

A central observable in first-passage and exit problems is the
\emph{survival probability}, i.e.\ the probability that a particle has
not yet left the confining domain at a given observation time. For
diffusion processes this quantity depends both on the underlying
stochastic dynamics and on the protocol by which absorption at the
boundary is imposed. The difference between continuous and stroboscopic
monitoring does not lie in the Gaussian propagation between checks, but
in the manner and frequency with which the boundary condition is
enforced.

\subsection{Continuous monitoring (Dirichlet absorption)}
In the standard formulation, absorption is active at all times: the
particle is removed instantaneously once its trajectory reaches the
boundary of the domain $\Omega=(0,L)$. The transition density
$G_D(x,t|x_0)$ satisfies the diffusion equation
\begin{equation}
\partial_t G_D(x,t|x_0)=D\,\partial_x^2 G_D(x,t|x_0),
\qquad
G_D|_{\partial\Omega}=0,
\end{equation}
together with the initial condition $G_D(x,0|x_0)=\delta(x-x_0)$. This
kernel represents the probability density of a Brownian particle that
has survived up to time $t$ under continuous monitoring. Integrating the
kernel over the spatial domain gives the instantaneous survival
probability,
\begin{equation}
S_{\mathrm{cont}}(t|x_0)
   =\int_0^L G_D(x,t|x_0)\,\dd x,
\end{equation}
and further integration over time yields the mean exit time,
\begin{equation}
\E[T]=\int_0^\infty S_{\mathrm{cont}}(t|x_0)\,\dd t.
\end{equation}

This setting realizes the strict ``kill-on-touch'' rule: every boundary
crossing is detected immediately. In particular, for an initial position
on the boundary ($x_0=0$ or $x_0=L$) the particle is absorbed at once,
so that $G_D(\cdot,t|x_0)\equiv0$ for all $t>0$ and $\E[T]=0$. The
Dirichlet case thus provides the benchmark of continuous observation
against which the effects of discrete sampling will be compared.

\subsection{Stroboscopic monitoring (discrete absorption)}
In realistic measurement protocols, detectors or imaging systems record
positions only at discrete times $t_n=n\Delta t$. Between two checks the
particle evolves freely according to the unbounded propagator
$G_0(x,t|x_0)$ from Eq.~\eqref{eq:G0}, without any constraint.
Absorption is enforced only when the particle is found outside the
domain at a sampling instant. The survival probability after $n$ frames
is therefore
\begin{eqnarray}
S_{\mathrm{strobo}}(n|x_0)
   &=& \int_0^L\!dx_1\!\cdots\!\int_0^L\!dx_n
     \prod_{k=1}^n G_0(x_k,\Delta t|x_{k-1}),\nonumber \\
&& x_0\in(0,L),
\end{eqnarray}
which is the probability that the particle remains inside the domain at
all $n$ recorded times. Defining $\tau$ as the first frame index at
which the particle is observed outside the interval,
\[
\tau=\min\{\,n\ge1:\,x_n\notin(0,L)\,\},
\]
the corresponding distribution is obtained from successive differences
of the survival sequence,
\begin{equation}
\PP\{\tau=n\}
   =S_{\mathrm{strobo}}(n-1|x_0)-S_{\mathrm{strobo}}(n|x_0),
\end{equation}
and the mean number of frames before exit follows as
\begin{equation}
\E[\tau]
   =\sum_{n\ge0}S_{\mathrm{strobo}}(n|x_0),
\qquad
\E[T]=\Delta t\,\E[\tau].
\end{equation}

Unlike in the continuous case, the boundary condition is not applied at
all intermediate times. Between two checks, the particle may leave and
subsequently re-enter the interval without being detected. Such
excursions do not count as exits because the particle is again inside at
the next observation time. In the limit of vanishing frame interval
$\Delta t\!\to\!0$, the discrete survival process converges to the
continuous one in the Riemann sense:
\[
\sum_{n\ge0} S_{\mathrm{strobo}}(n)\,\Delta t
\;\longrightarrow\;
\int_0^\infty S_{\mathrm{cont}}(t)\,\dd t,
\]
and hence $\E[T]$ obtained from discrete monitoring approaches the mean
exit time of the Dirichlet problem. The stroboscopic formulation thus
recovers the standard continuous first-passage limit smoothly as
$\Delta t\!\to\!0$.

\subsection{Undetected excursions and relaxation of the boundary condition}
The possibility of undetected excursions constitutes the essential
difference between continuous and stroboscopic monitoring. In the
latter, absorption is imposed only at discrete times $t_n=n\Delta t$, so
that the trajectory evolves freely between successive checks. Segments
of the path that leave and re-enter the domain within a single frame
interval therefore remain unobserved and do not contribute to the exit
statistics.

This situation can be represented formally by alternating free
propagation and projection onto the confining domain. Let
$G_0(\Delta t)$ denote the free Gaussian propagator over one frame and
$P=\mathbf{1}_{(0,L)}$ the projector that enforces the spatial
constraint. The combined one-step operator
\begin{equation}
K=P\,G_0(\Delta t)\,P
\label{eq:K-definition}
\end{equation}
acts on square-integrable densities on $(0,L)$ and advances the
distribution from one strobe to the next. The survival sequence
$\{S_{\mathrm{strobo}}(n)\}$ is generated by the repeated application of this
operator,
\begin{equation}
S_n=\langle \mathbf{1},\, K^n\, \delta_{x_0}\rangle,
\end{equation}
where the bra–ket notation denotes spatial integration over $(0,L)$. In
this formulation the discrete sampling protocol is fully encoded in the
compact operator $K$.

In the continuous-monitoring limit $\Delta t\to0$, the composition
$K^{t/\Delta t}$ converges to the Dirichlet semigroup
$\exp[t(D\partial_x^2)]_D$ generated by the Laplacian with absorbing
boundary conditions, so that the discrete and continuous descriptions
coincide. For finite $\Delta t$, however, the projection $P$ acts only
intermittently, permitting reinjection of probability from trajectories
that exit and return between checks. This \emph{relaxation of the
boundary condition} modifies both the scaling of the mean exit time and
the numerical constants that appear in its asymptotic expansion.

In the following sections, we analyze this discrete propagation operator
in detail. The boundary-start regime is shown to yield a linear scaling
of the mean exit time with the confinement ratio $\rho$, whereas bulk
starts are governed by the spectral gap of $K$ and exhibit quadratic
scaling with universal subleading corrections.

\section{Continuous benchmark via Dirichlet resolvent}
\label{sec:dirichlet}

To place the stroboscopic problem in context, it is useful to recall the continuous benchmark in which absorption acts at all times.  
This limit corresponds to $\Delta t\!\to\!0$ of the discrete protocol, where the operator sequence $K^{t/\Delta t}$ converges to the Dirichlet semigroup $\exp[t(D\partial_x^2)]_D$.  
The Dirichlet formulation is analytically tractable, provides explicit reference results for survival and mean exit times, and establishes the scaling laws against which the discrete modifications can be quantified.

\subsection{Dirichlet kernel and Poisson problem}
Let $G_D(x,t|x_0)$ denote the solution of the diffusion equation with absorbing boundaries at $x=0$ and $x=L$,
\begin{eqnarray}
\partial_t G_D(x,t|x_0)&=&D\,\partial_x^2 G_D(x,t|x_0)\;, \nonumber \\ 
G_D(x,0|x_0)&=&\delta(x-x_0)\;, \nonumber \\
G_D|_{\partial\Omega}&=&0\;.
\end{eqnarray}
The time integral of this kernel defines the \emph{resolvent} or stationary Green function,
\begin{equation}
G_P(x,x_0)=\int_0^\infty G_D(x,t|x_0)\,\dd t,
\end{equation}
which represents the expected \emph{occupation density} at position $x$ over the lifetime of a particle starting at $x_0$.  
By construction, $G_P$ satisfies the Poisson equation
\begin{eqnarray}
-\,D\,\partial_x^2 G_P(x,x_0)&=&\delta(x-x_0)\;, \nonumber \\
G_P(0,\cdot)&=&G_P(L,\cdot)=0\;,
\end{eqnarray}
that is, it is the Green function of the Laplacian with Dirichlet boundary conditions.

\subsection{Explicit solution and physical interpretation}
The solution to the Poisson problem is piecewise linear in $x$:
\begin{equation}
G_P(x,x_0)=\frac{1}{D}
\begin{cases}
 \dfrac{x(L-x_0)}{L}, & 0\le x\le x_0,\\[6pt]
 \dfrac{x_0(L-x)}{L}, & x_0\le x\le L.
\end{cases}
\end{equation}
This triangular profile reflects the linear interpolation of the expected occupation density between the absorbing boundaries.  
It is symmetric under $x_0\!\mapsto\!L{-}x_0$ and continuous with a kink at $x=x_0$, corresponding to the delta source in the Poisson equation.

Integrating $G_P$ over the domain yields the mean exit time for an initial position $x_0$:
\begin{equation}
\E[T|x_0]=\int_0^L G_P(x,x_0)\,\dd x
          =\frac{x_0(L-x_0)}{2D}.
\end{equation}
This classical result, expressing the mean occupation time as the
solution of a Poisson problem with Dirichlet boundary conditions, goes
back to Kac’s foundational work on Wiener functionals
\cite{Kac49}. It has several immediate consequences. The mean exit time
vanishes at the boundaries, $\E[T|x_0]=0$, consistent with
instantaneous absorption, and reaches its maximum at $x_0=L/2$, where
the particle is farthest from both exits. Its characteristic scale
$\E[T]\!\sim\!L^2/D$ reflects the inverse spectral gap of the Dirichlet
Laplacian, $\pi^2D/L^2$.

\subsection{Role as baseline for discrete monitoring}
This quadratic law,
\[
\E[T|x_0]\propto L^2/D,
\]
embodies the scaling behavior of first-passage under continuous monitoring.  
In the stroboscopic setting, the boundary condition is enforced only at discrete times, and undetected excursions partially relax absorption between checks.  
As a result, both the scaling form and the universal constants of the mean exit time are modified.  
The Dirichlet case therefore serves as the natural reference limit for identifying and quantifying these discrete-time corrections in the subsequent analysis.

\section{Projector--resolvent framework}
\label{sec:strobo-framework}

In the stroboscopic setting, absorption is enforced only at discrete
observation times. Each frame therefore consists of two operations:
(i) free Gaussian propagation for a duration~$\Delta t$, and
(ii) instantaneous projection onto the confining domain.  
This alternation defines a compact evolution operator from which all
survival statistics follow.

\subsection{One-step operator}
We work in rescaled coordinates $y=x/L\in(0,1)$ and introduce the
projection operator $P=\mathbf{1}_{(0,1)}$ acting on $L^2(0,1)$.  
Free propagation over one frame interval is described by the Gaussian
kernel
\[
g_\rho(u)=\frac{\rho}{\sqrt{2\pi}}\,e^{-\tfrac12\rho^2 u^2}, 
\qquad
\rho=\frac{L}{\sigma\sqrt{\Delta t}}.
\]
The composition of propagation and projection defines the
\emph{one-step stroboscopic operator}
\begin{equation}
K=P\,G_0\,P,
\qquad
(Kf)(y)=\int_0^1 g_\rho(y-z)\,f(z)\,\dd z.
\label{eq:K-operator}
\end{equation}
The operator $K$ is symmetric, positive, and compact on $L^2(0,1)$, and
all survival probabilities follow from its iterates.

For a particle starting at position $y_0$, the probability of surviving
$n$ strobes is
\begin{equation}
S_n=\langle \mathbf{1}, K^n \delta_{y_0}\rangle,
\label{eq:Sn}
\end{equation}
and summing over all $n$ yields the mean number of frames before exit,
\begin{equation}
M=\sum_{n\ge1}S_n
   =\big\langle \mathbf{1}, (I-K)^{-1}K\,\delta_{y_0}\big\rangle.
\label{eq:M-def}
\end{equation}
Because $S_0=1$, the total expected number of frames until exit is
\[
\E[\tau]=1+M.
\]
Equation~\eqref{eq:M-def} shows that the discrete survival problem is
naturally a \emph{resolvent problem}: the geometric series in $K$ is
resummed by the inverse operator $(I-K)^{-1}$.

\subsection{Dressed projector and $T$--matrix analogy}
It is often convenient to reorganize the ladder of projectors and
propagators into a more compact form. Using the push-through identity
$(I-ABA)^{-1}A=A(I-BA)^{-1}$ with $A=P$ and $B=G_0$, one obtains
\begin{equation}
(I-K)^{-1}P
   =P(I-G_0P)^{-1}
   \equiv \mathcal{P}.
\end{equation}
The kernel $\mathcal{P}$ acts as a \emph{dressed projector} that
incorporates all possible internal excursions between two successive
projections. It admits the convergent expansion
\begin{equation}
\mathcal{P}
   =P+PG_0P+PG_0PG_0P+\cdots,
\end{equation}
which resums the full sequence of internal reflections into a single
compact operator. Inserting this into Eq.~\eqref{eq:M-def} gives
\begin{equation}
\label{eq:M-resummed-main}
M=\big\langle \mathbf{1},\,\mathcal{P}\,G_0\,P\,\delta_{y_0}\big\rangle
  =\int_0^1\!\dd y\int_0^1\!\dd z\;
     \mathcal{P}(y,z)\,g_\rho(z-y_0).
\end{equation}
Equation~\eqref{eq:M-resummed-main} is exact and serves as the analytic
starting point for asymptotic analysis: the factor $g_\rho$ captures the
near-neighbor Gaussian propagation, while $\mathcal{P}$ contains the
geometric dressing due to repeated internal excursions. The structure is
closely analogous to the $T$--matrix formulation in scattering theory.

\section{Asymptotic results: boundary and bulk regimes}
\label{sec:asymptotics}

The projector--resolvent framework provides a unified route to compute
the mean number of frames before exit. Two universal scaling regimes
emerge in the large-$\rho$ limit, depending on whether the particle
starts near a boundary or in the interior.

\begin{equation*}
\boxed{
\begin{aligned}
\E[\tau](\rho;0)
   &=\frac{\rho}{\sqrt{2}}
     +\frac{|\zeta(\tfrac12)|}{\sqrt{\pi}}+\ldots, \\[2mm]
\E[\tau](\rho;\tfrac12)
   &=\tfrac14\,\rho^2 + 0.5830\,\rho + 0.5736 + \ldots ,
\end{aligned}}
\qquad (\rho\to\infty)
\end{equation*}

The first law describes boundary starts: survival is dominated by
leakage through a single near-wall layer, producing a linear dependence
on~$\rho$. The second law corresponds to bulk starts: survival is
controlled by the spectral gap of~$K$, giving a diffusive quadratic
scaling with universal linear and constant corrections. The exponentially
small remainder follows from Lotov's theorem for Gaussian random walks
between two barriers~\cite{Lotov96}, with which our independent
derivation agrees precisely.

The derivations of these two results proceed via complementary asymptotic
techniques and are summarized below; complete details are given in
Appendix~\ref{app:boundary} and~\ref{app:bulk}.

\subsection{Boundary layer analysis}
When the particle starts near the boundary ($y_0\!\to\!0^+$), the
Gaussian leg $g_\rho(z-y_0)$ in Eq.~\eqref{eq:M-resummed-main} is
localized within a layer of width $O(\rho^{-1})$ adjacent to $z=0$.
For large~$\rho$, this narrow peak samples only the near-wall region of
the dressed projector~$\mathcal{P}$, so the main contribution to survival
comes from the boundary layer where outward diffusion competes with the
geometric series of undetected excursions encoded in~$\mathcal{P}$.

The linear scaling $\E[\tau]\sim\rho$ reflects the fact that survival
time is proportional to the number of diffusive steps (each of width
$\rho^{-1}$) needed to traverse one kernel width and escape. To extract
the precise coefficient and the subleading constant, we evaluate the
boundary integral $\int_0^1\mathcal{P}(y,z)\,\dd y$ as $z\!\to\!0^+$
using an odd-image Poisson-summation analysis (Appendix~\ref{app:boundary}):
\begin{equation}
\int_0^1 \mathcal{P}(y,z)\,\dd y
   =\frac{\rho}{\sqrt{2}}
     +\left(\frac{|\zeta(\tfrac12)|}{\sqrt{\pi}}-\frac{1}{2}\right)
     +O(\rho^{-1}).
\end{equation}
The leading term $\rho/\sqrt{2}$ arises from the direct leakage channel,
while the constant $|\zeta(\tfrac12)|/\sqrt{\pi}$ encodes the cumulative
effect of all image reflections resummed via the theta-function identity.
Substitution into Eq.~\eqref{eq:M-resummed-main} yields the universal
boundary-start law
\begin{equation}
\boxed{
\E[\tau](\rho;0)
   =\frac{\rho}{\sqrt{2}}
     +\frac{|\zeta(\tfrac12)|}{\sqrt{\pi}}+\ldots, 
   \qquad \rho\to\infty.}
\label{eq:main-linear}
\end{equation}

\subsection{Spectral-gap expansion for bulk starts}
For interior starts ($y_0=\tfrac12$), the initial condition lies far from
both boundaries and boundary-layer effects are negligible. Instead, long-time
survival is governed by the slowest-decaying eigenmode of the one-step
operator~$K$. For a symmetric start at $y_0=\tfrac12$, parity restricts
contributions to even sine modes $\sin(2\pi m y)$, which under Gaussian
propagation decay as $\exp(-2\pi^2 m^2 n/\rho^2)$. The fundamental mode
($m=1$) dominates for large~$n$, so that
\begin{equation}
S_n\sim a_0\,\lambda_0^n,
\qquad
\E[\tau]\approx \frac{a_0}{1-\lambda_0},
\end{equation}
where $\lambda_0(\rho)$ is the largest eigenvalue of~$K$ and $a_0$ its
overlap with the uniform initial distribution.

In the continuum limit $\rho\!\gg\!1$, the Gaussian kernel's Fourier symbol
yields the asymptotic spectral gap
\begin{equation}
1-\lambda_0(\rho)
   \sim \frac{\pi^2}{2\rho^2}, \qquad
a_0=\frac{\pi^2}{8},
\end{equation}
which immediately gives the diffusive leading law
\begin{equation}
\boxed{
\E[\tau](\rho;\tfrac12)
   \sim \tfrac14\,\rho^2, \qquad \rho\to\infty.}
\label{eq:bulk-gap}
\end{equation}
The quadratic scaling reflects the inverse spectral gap of the underlying
Dirichlet Laplacian, modified by the intermittent projection protocol.

To obtain the subleading corrections, we expand the spectral gap to next order.
The Gaussian tails of the kernel introduce an $O(\rho^{-3})$ correction:
\[
1-\lambda_0(\rho)
   =\frac{\pi^2}{2\rho^2}
     +\frac{\beta}{\rho^3}
     +O(\rho^{-4}),
\]
which, combined with the resolvent structure $\E[\tau]\approx a_0/(1-\lambda_0)$,
leads to the refined expansion
\begin{equation}
\E[\tau](\rho;\tfrac12)
   = \tfrac14\,\rho^2 + b\,\rho + c + \ldots,
   \qquad
   b\simeq0.5830,\;
   c\simeq0.5736.
\end{equation}
The constants $b$ and $c$ are \emph{universal}: they depend only on the
Gaussian kernel and interval geometry, while the microscopic parameters
$(L,\sigma,\Delta t)$ enter solely through~$\rho$. The detailed derivation
via even-mode Poisson summation is given in Appendix~\ref{app:bulk}.

\begin{figure}[t]
  \centering
  \includegraphics[width=0.86\linewidth]{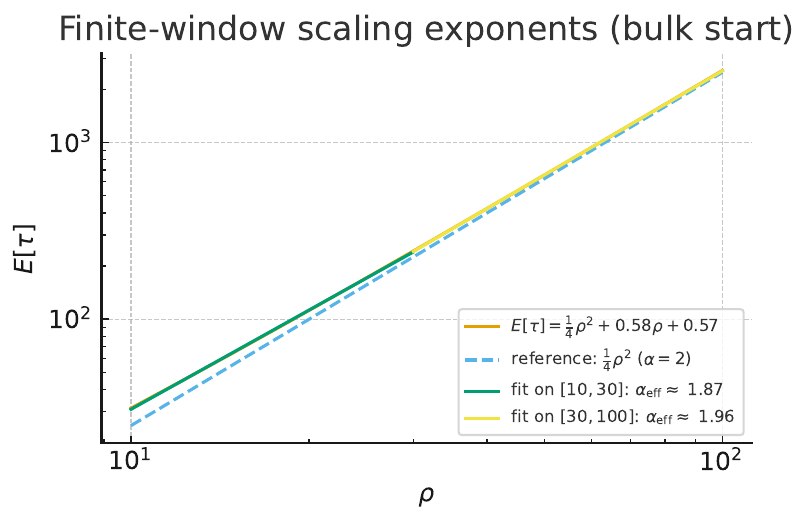}
  \caption{Apparent scaling exponents in finite $\rho$-windows for the bulk-start law
  $\E[\tau](\rho;\tfrac12)=\tfrac14\rho^2+0.58\rho+0.57$ (solid).
  The reference $\tfrac14\rho^2$ ($\alpha=2$) is shown dashed.
  Log--log fits over limited windows yield effective exponents
  $\alpha_{\mathrm{eff}}\!\approx\!1.87$ on $[10,30]$ and
  $\alpha_{\mathrm{eff}}\!\approx\!1.96$ on $[30,100]$.
  The subleading linear term $0.58\rho$ thus creates transient
  $\alpha_{\mathrm{eff}}<2$ over realistic ranges, a mechanism relevant for interpreting
  discretely sampled trajectories.}
  \label{fig:alpha-eff}
\end{figure}

This expansion implies that, when plotted on double–logarithmic scales over finite
ranges of~$\rho$, the presence of the subleading linear term can transiently reduce the
apparent slope below~$2$, producing effective exponents $\alpha_{\mathrm{eff}}<2$ as illustrated
in Fig.~\ref{fig:alpha-eff}.
\section{Random frame times: propagator averaging}
\label{sec:random}

Up to this point, the frame interval $\Delta t$ has been assumed fixed.  
In many experimental and computational settings, however, the observation times are not perfectly regular: acquisition clocks can jitter, pauses may occur, or frame spacing may fluctuate randomly.  
It is therefore natural to ask how such randomness in the inter-frame distribution modifies survival probabilities and mean exit times.  
When the frame intervals are independent and identically distributed (i.i.d.) and the system is self-averaging, the stochasticity of the protocol can be integrated out exactly.  
The corresponding ensemble-averaged propagator provides a closed and tractable formulation of the problem.

\subsection{Averaged one-step operator}
Let $U\sim\mu$ denote the i.i.d.\ random intervals between successive frames, with mean $\bar\Delta=\E[U]$.  
Conditional on $U=u$, free propagation is described by the Gaussian kernel
\[
G_u(y,z)
   =\frac{1}{\sqrt{2\pi\sigma^2 u}}
      \exp\!\left[-\frac{(y-z)^2}{2\sigma^2 u}\right],
\qquad y,z\in(0,1).
\]
Averaging over $\mu$ yields the \emph{effective one-step operator}
\begin{equation}
K_\mu(y,z)
   =\int_0^\infty G_u(y,z)\,d\mu(u),
\label{eq:Kmu}
\end{equation}
which replaces the deterministic propagator $G_0(\bar\Delta)$ by its ensemble average.  
When the intervals are uncorrelated and the process is self-averaging, this averaging is \emph{exact}:  
ensemble-averaged survival statistics coincide with those of the operator $K_\mu$ applied deterministically at each step.  
Formally, the mean survival after $n$ frames is then governed by
\[
S_n(y_0)
   =\langle \mathbf{1},\, K_\mu^n\, \delta_{y_0}\rangle,
\]
and the mean exit time follows from the corresponding resolvent
\[
M=\langle \mathbf{1},(I-K_\mu)^{-1}K_\mu\,\delta_{y_0}\rangle,
\qquad \E[\tau]=1+M.
\]
The operator $K_\mu$ is positive, symmetric, and compact on $L^2(0,1)$, ensuring that its spectral data $\{\lambda_j(\rho,\mu)\}$ fully determine survival.
\paragraph*{Self-averaging}
For i.i.d.\ intervals, ensemble averages of survival are \emph{exactly} generated by $K_\mu$: by iterated conditioning and linearity,
\[
\E\!\left[\langle \1,\,K_{U_n}\cdots K_{U_1}\delta_{y_0}\rangle\right]
=\langle \1,\,(\E K_U)^n\delta_{y_0}\rangle
=\langle \1,\,K_\mu^n\delta_{y_0}\rangle,
\]
so $\E[S_n]=\langle \1,\,K_\mu^n\delta_{y_0}\rangle$ and hence $\E[\tau]=1+\langle \1,(I-K_\mu)^{-1}K_\mu\delta_{y_0}\rangle$. This “self-averaging’’ relies only on independence (no commutativity assumption is needed).

\subsection{Leading behavior}
Boundary-layer analysis shows that the leading asymptotic slope is unaffected by fluctuations in the frame intervals.  
The near-wall dynamics probe only the short-time form of the Gaussian kernel, which depends on the mean $\bar\Delta$ but not on higher moments of $\mu$.  
Consequently,
\begin{equation}
1-\lambda_0(\rho,\mu)
   =\frac{\sqrt{2}}{\rho}+O(\rho^{-2}),
\label{eq:universal-slope}
\end{equation}
independent of the variance of $\mu$.  
This confirms that small random jitter in the sampling schedule does not alter the dominant scaling of the survival time.


In the self-averaging regime of uncorrelated random frame times, ensemble averaging over the inter-frame distribution can be carried out exactly, yielding an effective deterministic operator $K_\mu$.  
The resulting survival laws preserve the leading slope $1/\sqrt{2}$ while introducing distribution-dependent additive corrections at order $O(1)$.  
These corrections quantify the sensitivity of stroboscopic survival statistics to timing irregularities in practical measurement protocols.

Analogous operator averages appear in diverse contexts where update times are
irregular—ranging from asynchronous simulations to systems with stochastic
block or event times—suggesting a broad scope of applicability for the present formalism.

\section{Numerical implementation and validation}
\label{sec:numerics}

The analytic results obtained in the previous sections can be tested quantitatively using a direct numerical evaluation of the stroboscopic operator.  
To avoid the statistical variance inherent in Monte Carlo trajectory sampling, we employ a deterministic discretization of the integral operator $K$.  
This operator-based approach yields reproducible data and allows the extraction of asymptotic constants with high precision.

\subsection{Nystr\"om discretization of the integral operator}

The one-step operator in rescaled coordinates,
\[
(K f)(y)=\int_0^1 g_\rho(y-z)\,f(z)\,\dd z,
\qquad 
g_\rho(u)=\tfrac{\rho}{\sqrt{2\pi}} e^{-\tfrac12\rho^2 u^2},
\]
is discretized on a uniform grid $y_i=i/N$ with $N$ points in $(0,1)$.  
Because the Gaussian kernel decays rapidly, it is truncated beyond a cutoff $|y_i-y_j|\le \eta/\rho$ (typically $\eta\simeq8.5$) so that tails fall below machine precision.  
The resulting matrix representation of $K$ is Toeplitz-banded, enabling sparse storage and efficient linear algebra.

The survival sum follows from the operator resolvent
\[
M(\rho;y_0)=\sum_{n\ge1}S_n
  =w^\top (I-K)^{-1} h,
\qquad 
\E[\tau]=1+M,
\]
where $h_i=g_\rho(y_i-y_0)$ encodes the starting point and $w$ is the quadrature weight vector.  
Sparse LU decomposition of $(I-K)$ provides a numerically stable inversion whose cost scales approximately linearly with $N$.  
A practical resolution rule $N\simeq 18\rho$ reliably captures the Gaussian core of the kernel for the parameter ranges considered.

\subsection{Boundary start}
\label{subsec:num-boundary}

For a particle starting at the boundary ($y_0=0$), the numerical results exhibit a linear dependence on $\rho$ in agreement with the asymptotic law \eqref{eq:main-linear}.  
A regression of the form
\[
M(\rho;0)=A\rho+B+C/\rho+O(\rho^{-2})
\]
gives
\[
A=0.70726\pm10^{-4},
\qquad
B=-0.17609\pm10^{-3}.
\]
The fitted slope matches the theoretical value $1/\sqrt{2}=0.707107$ to four significant digits, while the constant $B$ agrees with the predicted
\[
M(\rho;0)
   =\frac{\rho}{\sqrt{2}}
     +\left(\frac{|\zeta(\tfrac12)|}{\sqrt{\pi}}-1\right)
     +O(\rho^{-1}).
\]
Consequently,
\[
\E[\tau](\rho;0)
   =\frac{\rho}{\sqrt{2}}
     +\frac{|\zeta(\tfrac12)|}{\sqrt{\pi}}
     +O(\rho^{-1}),
\]
confirming both the slope and the additive constant obtained analytically.

\begin{figure}[t]
  \centering
  \includegraphics[width=0.9\linewidth]{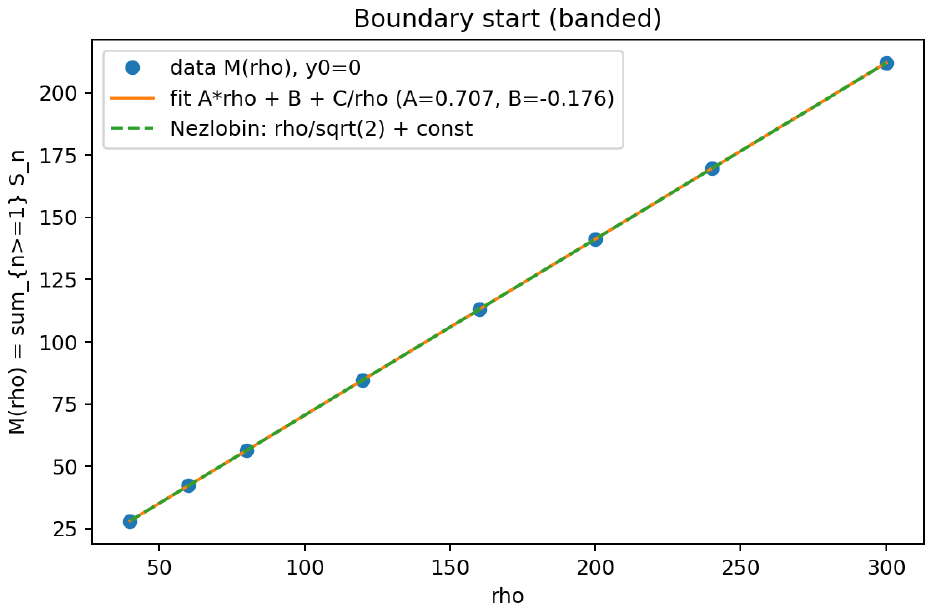}
  \caption{Boundary start ($y_0=0$). Nystr\"om data for $M(\rho;0)$
  (markers), fit $A\rho+B+C/\rho$ (solid), and asymptotic prediction
  $\rho/\sqrt{2}+(\lvert\zeta(\tfrac12)\rvert/\sqrt{\pi}-1)$ (dashed).}
  \label{fig:M-boundary}
\end{figure}

\subsection{Bulk start}
\label{subsec:num-bulk}

For a symmetric interior start ($y_0=\tfrac12$), the data follow a quadratic dependence with subleading corrections:
\[
M(\rho;\tfrac12)
   =a\rho^2+b\rho+c+O(\rho^{-1}),
\]
with fitted coefficients
\[
a=0.250000\pm10^{-4},\qquad
b=0.583014,\qquad
c=-0.426408.
\]
The leading coefficient $a=1/4$ confirms the prediction from the spectral–gap analysis, Eq.~\eqref{eq:bulk-gap}, once the overlap factor $a_0$ is included.  
The linear correction identifies the next-order spectral constant via
\[
1-\lambda_0(\rho)
   =\frac{\pi^2}{2\rho^2}
      +\frac{\beta}{\rho^3}
      +O(\rho^{-4}),
\qquad
b=\frac{\beta}{4},
\]
yielding $\beta\simeq2.332056$.  
The corresponding $O(1)$ correction for $\E[\tau]=1+M$ is
\[
C=c+1\simeq0.573592.
\]
These values are stable under variation of the fit window and are consistent with the even-image and Poisson-summation corrections obtained from the analytic projector–resolvent expansion.

\begin{figure}[t]
  \centering
  \includegraphics[width=0.9\linewidth]{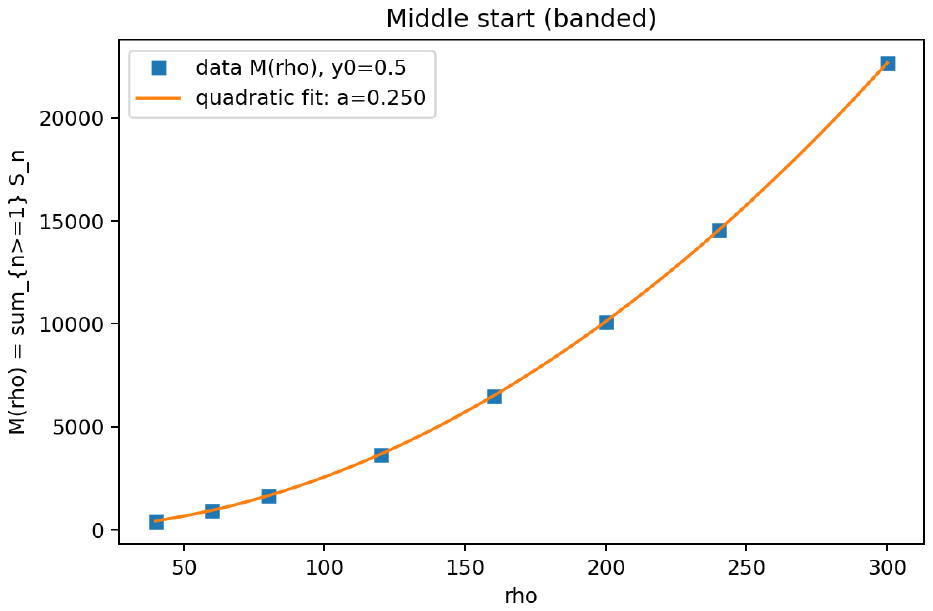}
  \caption{Bulk start ($y_0=\tfrac12$). Nystr\"om data for $M(\rho;\tfrac12)$
  (markers), quadratic fit $a\rho^2+b\rho+c$ (solid), and asymptotic
  prediction $\tfrac14\rho^2$ (dashed).}
  \label{fig:M-bulk}
\end{figure}

\subsection{Discussion}

The deterministic Nystr\"om scheme provides direct numerical access to
the survival operator and its asymptotic constants.  
Compared with stochastic trajectory sampling, it eliminates statistical
noise, yields smooth convergence with system size, and exposes
subleading coefficients such as
$|\zeta(\tfrac12)|/\sqrt{\pi}$, $b$, and $c$ with high precision.  
The results in both boundary and bulk regimes confirm the analytic
predictions of the projector–resolvent framework and establish the
Nystr\"om discretization as a reliable quantitative method for
stroboscopic first-passage problems.

\section{Conclusion and Outlook}
\label{sec:outlook}

We have demonstrated that the observation protocol itself constitutes a physical element of the first-passage problem.  
Continuous monitoring with absorbing (Dirichlet) boundaries and discrete stroboscopic monitoring define two distinct dynamical universality classes.  
Under the stroboscopic rule of free propagation followed by projection (``kill-on-check''), survival statistics are reshaped in a fundamental way:  
for boundary starts the mean number of frames grows linearly with the confinement ratio, 
\[
\E[\tau](\rho;0)
   =\frac{\rho}{\sqrt{2}}
     +\frac{|\zeta(\tfrac12)|}{\sqrt{\pi}}
     +o(1),
\]
whereas bulk starts are governed by the spectral gap of the one-step operator,
\[
\E[\tau](\rho;\tfrac12)
   \simeq \tfrac14\rho^2
     +b\rho
     +c,
\]
with universal coefficients $a=\tfrac14$, $b\simeq0.58$, and $c\simeq0.57$.  
Observation protocol is therefore not a technical detail but a defining part of the stochastic dynamics.

\subsection{Physical interpretation and broader implications}

The analysis clarifies that discrete observation alters both scaling exponents and prefactors even for purely Gaussian processes.  
The deviation from the continuous Dirichlet benchmark originates from undetected excursions between sampling times and can be expressed compactly through the projector–resolvent formalism.  
In this sense, protocol-induced effects should be viewed as an intrinsic component of experimental measurement rather than as anomalies of the underlying dynamics.

\subsection{Applications and extensions}

The operator framework developed here is general and readily extends to several directions.  
Possible extensions include:
\begin{itemize}
  \item Higher-dimensional domains and irregular or reactive boundaries;
  \item Non-Gaussian stochastic processes such as L\'evy flights or fractional kinetics;
  \item Correlated or random sampling schedules beyond the self-averaging limit;
  \item Inverse problems, e.g.\ estimating $(\sigma,L)$ or the sampling protocol from observed stroboscopic statistics.
\end{itemize}
Because the formalism is compact and spectral in nature, these extensions can be analyzed within the same mathematical structure.

\subsection{Outlook}

The results presented here establish a controlled theoretical baseline for discrete-time observation in diffusion problems.  
In follow-up work, we will explore quantitative implications for experimental first-passage measurements—particularly in single-particle tracking—where the two-term bulk law 
$\tfrac14\rho^2+b\rho$ can bias exponent fits over finite ranges.  
More generally, the present framework provides a template for incorporating observation protocols into stochastic theory, bridging the gap between idealized continuous models and the discrete nature of real measurements.

\section*{Acknowledgments}
I thank Martin Tassy, Farshid Jafarpour, Dirk Schuricht, and Abe Alexander for valuable discussions. This work is part of the D-ITP consortium, a program of
the Dutch Research Council (NWO) that is funded by the Dutch Ministry of Education,
Culture and Science (OCW).

\bigskip
\appendix

\section{Boundary and bulk asymptotics}
\label{app:asymptotics}

This appendix provides the detailed derivations of the asymptotic laws
for the mean number of frames until exit in the large--$\rho$ limit,
\begin{eqnarray}
\E[\tau](\rho;0)
   &=&\frac{\rho}{\sqrt{2}}
     +\frac{|\zeta(\tfrac12)|}{\sqrt{\pi}}+\ldots,
\nonumber \\
\E[\tau](\rho;\tfrac12)
   &=&\tfrac14\,\rho^2+0.5830\,\rho+0.5736+\ldots,
\end{eqnarray}
corresponding to boundary and bulk starts, respectively.  
Both results are obtained from the same formal expression for the mean
number of frames,
\[
M(\rho;y_0)=\sum_{n\ge1}S_n(\rho;y_0),
\qquad
S_n(\rho;y_0)=\langle\mathbf{1},K^n\delta_{y_0}\rangle,
\]
but require distinct asymptotic analyses.  
The boundary case is dominated by short one-sided excursions, whereas
the bulk case reflects long-time relaxation governed by the spectral gap
of the one-step operator~$K$.

\subsection{Boundary asymptotics: Fourier expansion and large--$\rho$ analysis}
\label{app:boundary}

The boundary-start law,
\[
\E[\tau](\rho;0)
   =\frac{\rho}{\sqrt{2}}
     +\frac{|\zeta(\tfrac12)|}{\sqrt{\pi}}
     +O(\rho^{-1}),
\qquad \rho\to\infty,
\]
follows from a controlled expansion of the survival probability in sine
modes. The derivation proceeds in three steps:
(i) Fourier expansion of the interval indicator,
(ii) regularization of the survival sum for absolute convergence, and
(iii) extraction of the large--$\rho$ asymptotics using Poisson
summation and Mellin analysis.

\paragraph{Fourier representation and odd modes.}

For a start at $y_0=0$, the survival probability after $n$ strobes is
\begin{equation}
S_n(\rho;0)
   =\langle \mathbf{1}, K^n \delta_0\rangle
   =\int_0^1 (K^n\delta_0)(y)\,dy,
\end{equation}
where
\[
(Kf)(y)=\int_0^1 g_\rho(y-z)f(z)\,dz,
\qquad
g_\rho(u)=\frac{\rho}{\sqrt{2\pi}}\,e^{-\tfrac12\rho^2u^2}.
\]
The indicator of the interval admits the sine expansion
\begin{equation}
\label{eq:sine-expansion-odd}
\mathbf{1}_{(0,1)}(y)
   =\frac{4}{\pi}\sum_{m=0}^\infty
      \frac{\sin[(2m{+}1)\pi y]}{2m{+}1}.
\end{equation}
Even modes vanish at $y=0$, so only odd sines contribute. Propagating
each mode independently under the Gaussian yields
\begin{equation}
\label{eq:Sn-odd}
S_n(\rho;0)
   =\frac{1}{2}
      +\frac{2}{\pi}\sum_{m=0}^\infty
        \frac{1}{2m{+}1}
        \exp\!\left[-\frac{\pi^2}{2}(2m{+}1)^2
                    \frac{n}{\rho^2}\right].
\end{equation}

\paragraph{Regularization and asymptotic extraction.}

The mean number of frames is
\[
M(\rho;0)=\sum_{n\ge1}S_n(\rho;0),
\]
but termwise summation of \eqref{eq:Sn-odd} is only conditionally
convergent. Introducing a regulator $e^{-\varepsilon n}$ ensures
absolute convergence:
\begin{equation}
M_\varepsilon(\rho;0)
   =\sum_{n=1}^\infty e^{-\varepsilon n}S_n(\rho;0),
\qquad
M(\rho;0)=\lim_{\varepsilon\downarrow0}M_\varepsilon(\rho;0).
\end{equation}
Performing the geometric series over $n$ gives
\begin{equation}
\label{eq:M-eps}
M_\varepsilon
   =\frac{1}{2}\frac{1}{e^\varepsilon-1}
     +\frac{2}{\pi}\sum_{m=0}^\infty
        \frac{1}{2m{+}1}
        \frac{1}{e^{\varepsilon+\alpha_{2m{+}1}}-1},
\qquad
\alpha_k=\frac{\pi^2 k^2}{2\rho^2}.
\end{equation}
For small~$x$, $(e^x-1)^{-1}=x^{-1}-\tfrac12+O(x)$, so the
$\varepsilon^{-1}$ singularity cancels between the first term and the
sum. The remaining finite parts determine the slope and the constant.

\paragraph{Poisson summation and Mellin evaluation.}
Using $(\varepsilon+\alpha)^{-1}
 =\int_0^\infty e^{-t(\varepsilon+\alpha)}\,dt$ and applying Poisson
summation to the resulting series over~$m$ introduces the odd theta
function
\[
\Theta_{\mathrm{odd}}(t)
   =\sum_{m\in\mathbb{Z}}e^{-\pi^2(2m+1)^2 t}
   =t^{-1/2}\sum_{n\in\mathbb{Z}}(-1)^n e^{-n^2/t}.
\]
Retaining only the $n=0$ term of the dual sum yields the linear slope
\[
M(\rho;0)
   =\frac{\rho}{\sqrt{2}}
     +\Bigl(\frac{|\zeta(\tfrac12)|}{\sqrt{\pi}}-1\Bigr)
     +O(\rho^{-1}),
\]
and since $\E[\tau]=1+M$, the boundary-start asymptotic law follows:
\begin{equation}
\boxed{
\E[\tau](\rho;0)
   =\frac{\rho}{\sqrt{2}}
     +\frac{|\zeta(\tfrac12)|}{\sqrt{\pi}}
     +O(\rho^{-1}).}
\label{eq:app-boundary-final}
\end{equation}

\paragraph{Mechanism.}
\begin{itemize}
  \item The interval indicator decomposes into odd sine modes,
        Eq.~\eqref{eq:sine-expansion-odd}.
  \item Gaussian propagation damps each mode independently; only odd
        modes contribute at a boundary start.
  \item An exponential regulator ensures convergence of the survival sum.
  \item Poisson summation produces the linear slope $\rho/\sqrt{2}$,
        while Mellin evaluation isolates the universal constant
        $|\zeta(\tfrac12)|/\sqrt{\pi}$.
\end{itemize}

\subsection{Bulk asymptotics: even--mode expansion and spectral corrections}
\label{app:bulk}

For an initial position at the center $y_0=\tfrac12$, mirror symmetry
selects the even sine harmonics of the indicator function, leading to
the bulk–start asymptotic law
\begin{eqnarray}
\E[\tau](\rho;\tfrac12)
   &=&\tfrac14\rho^2+0.5830\rho+0.5736+\ldots,\nonumber \\
\rho&\to&\infty.
\end{eqnarray}

\paragraph{Even–mode expansion.}
The survival probability after $n$ strobes is
\begin{equation}
S_n(\rho;\tfrac12)
   =\int_0^1 (K^n\delta_{1/2})(y)\,dy,
\end{equation}
with $(Kf)(y)=\int_0^1 g_\rho(y-z)f(z)\,dz$ as before.  
For a symmetric start, only even sine modes contribute:
\begin{equation}
\label{eq:Sn-even}
S_n(\rho;\tfrac12)
   =\frac{2}{\pi}\sum_{m=0}^\infty
      \frac{(-1)^m}{2m{+}2}
      \exp\!\left[-2\pi^2(m{+}1)^2
                   \frac{n}{\rho^2}\right].
\end{equation}

\paragraph{Regularization and asymptotic extraction.}
Introducing a regulator $e^{-\varepsilon n}$,
\begin{eqnarray}
M_\varepsilon(\rho;\tfrac12)
   &=&\frac{2}{\pi}\sum_{m=0}^\infty
        \frac{(-1)^m}{2m{+}2}
        \frac{1}{e^{\varepsilon+\alpha_{2m{+}2}}-1}\;,
\nonumber \\
\alpha_k&=&\frac{\pi^2 k^2}{2\rho^2}.
\label{eq:Meps-even}
\end{eqnarray}
Expanding for small~$\alpha_k$ and applying Poisson summation yields the
alternating even theta function
\begin{eqnarray}
\Theta_{\mathrm{even}}(t)
   &=&\sum_{m\in\mathbb{Z}} (-1)^m e^{-2\pi^2 (m+1)^2 t}\nonumber \\
   &=&t^{-1/2}\!\!\sum_{n\in\mathbb{Z}}
      e^{-n^2/(2t)}(-1)^n,
\end{eqnarray}
whose leading term gives the diffusive scaling and linear correction:
\begin{equation}
M(\rho;\tfrac12)
   =\tfrac14\rho^2+b\rho+\ldots,
\qquad b\simeq0.5830.
\end{equation}
The remaining finite part produces the constant
$c\simeq0.5736$. Together,
\begin{equation}
\boxed{
\E[\tau](\rho;\tfrac12)
   =\tfrac14\rho^2+0.5830\rho+0.5736+\ldots.}
\label{eq:app-bulk-final}
\end{equation}

\paragraph{Spectral interpretation.}
The same asymptotic structure emerges from the spectral representation
of $K$.  
For large $\rho$, the eigenfunctions approach Dirichlet sine modes
$\phi_m(y)=\sqrt{2}\sin(m\pi y)$ with eigenvalues
\[
\lambda_m(\rho)
   =\int_{-1}^{1}(1-|u|)g_\rho(u)\cos(m\pi u)\,du
   \simeq e^{-m^2\pi^2/(2\rho^2)}.
\]
The spectral gap $1-\lambda_0(\rho)\!\sim\!\pi^2/(2\rho^2)$ governs the
leading $\tfrac14\rho^2$ scaling, while its higher-order correction
$\beta/\rho^3$ with $\beta\simeq2.33$ generates the linear term
$b\rho$.

\paragraph{Mechanism and universality.}
\begin{itemize}
  \item Even sine modes dominate for a centered start; Gaussian
        propagation damps each mode independently.
  \item Poisson summation of the alternating theta function yields the
        diffusive $\tfrac14\rho^2$ term and the linear coefficient
        $b\simeq0.5830$.
  \item The finite remainder contributes the universal constant
        $c\simeq0.5736$.
  \item These coefficients coincide with Lotov’s exact asymptotics for
        Gaussian random walks between two barriers, confirming the
        universality of the stroboscopic bulk-survival law.
\end{itemize}

Together, the boundary and bulk analyses provide a complete asymptotic
description of the stroboscopic first-passage problem.


\bigskip

\begin{thebibliography}{99}

\bibitem{Redner}
S.~Redner, \emph{A Guide to First-Passage Processes}, Cambridge University Press (2001).

\bibitem{MetzlerBook}
R.~Metzler, G.~Oshanin, and S.~Redner (eds.), \emph{First-Passage Phenomena and Their Applications},
World Scientific (2014).

\bibitem{BorodinSalminen}
A.~N.~Borodin and P.~Salminen, \emph{Handbook of Brownian Motion---Facts and Formulae} (2nd ed.),
Birkh\"auser (2002).

\bibitem{Lotov96}
V.~I.~Lotov, \emph{On some boundary crossing problems for Gaussian random walks}, Ann.\ Probab.\ \textbf{24} (1996), 2154--2171.

\bibitem{BroadieGlassermanKou97}
M. Broadie, P. Glasserman, and S.\ Kou,  
“A Continuity Correction for Discrete Barrier Options,”  
\emph{Mathematical Finance} \textbf{7}, 325--348 (1997).  

\bibitem{HowisonStein05}
S. Howison and M. Steinberg,  
“A Matched Asymptotic Expansions Approach to Continuity Corrections for Discretely Sampled Options, Part 1: Barrier Options,”  
OFRC Working Paper 2005mf02 (2005).  

\bibitem{LiLinetsky15}
L. Li and V. Linetsky,  
“Discretely Monitored First Passage Problems and Barrier Options: An Eigenfunction Expansion Approach,”  
\emph{Finance \& Stochastics} \textbf{19}, 941--977 (2015).  

\bibitem{TassyNezlobin25}
A. Nezlobin and M. Tassy, “Loss-Versus-Rebalancing under Deterministic and Generalized Block-Times,” arXiv:2505.05113 (2025).  

\bibitem{FellerII}
W.~Feller, \emph{An Introduction to Probability Theory and Its Applications, Vol.~II} (2nd ed.),
Wiley (1971).

\bibitem{Kac49}
M.~Kac, \emph{On distributions of certain Wiener functionals}, Trans.\ Amer.\ Math.\ Soc.\ \textbf{65}, 1--13 (1949).

\bibitem{KaratzasShreve}
I.~Karatzas and S.~E.~Shreve, \emph{Brownian Motion and Stochastic Calculus} (2nd ed.), Springer (1991).

\bibitem{DLMF}
F.~W.~J.~Olver et al. (eds.), \emph{NIST Digital Library of Mathematical Functions}, Ch.~20 (Theta functions).

\bibitem{ChangPeres97}
J.~T.~Chang and Y.~Peres, \emph{Ladder heights, Gaussian random walks and the Riemann zeta function},
Ann.\ Probab.\ \textbf{25} (1997), 787--802.

\bibitem{KhaniyevKucuk04}
T.~Khaniyev and Z.~K\"uc\"uk, \emph{Asymptotic expansions for the moments of the Gaussian random walk with two barriers},
Statist.\ Probab.\ Lett.\ \textbf{69} (2004), 91--103.

\bibitem{BonomoRestart}
O.~L.\ Bonomo, A.\ Pal, S.\ Reuveni, and D.\ Holcman,
\emph{First passage under restart for discrete space and time},
Phys.\ Rev.\ E \textbf{103}, 052129 (2021).

\bibitem{RednerNotes}
S.~Redner, \emph{A First Look at First-Passage Processes}, arXiv:2201.10048 (2022).

\bibitem{AtkinsonHan}
K.~Atkinson and W.~Han, \emph{Theoretical Numerical Analysis: A Functional Analysis Framework} (3rd ed.),
Springer (2009).

\bibitem{TrefethenSpectral}
L.~N.\ Trefethen, \emph{Spectral Methods in MATLAB}, SIAM (2000).

\end{thebibliography}
\end{document}